\documentclass[review]{elsarticle}

\usepackage{lineno}
\usepackage[colorlinks=true,linkcolor=black,citecolor=blue,urlcolor=blue,pdfauthor=author]{hyperref}
\modulolinenumbers[5]
\usepackage{amsmath,amsfonts,amssymb,amsthm,bm}
\usepackage{float}
\usepackage{graphics}
\usepackage{graphicx}
\usepackage{caption}
\graphicspath{{figures/}}
\usepackage{subfigure}
\usepackage{fullpage}
\usepackage{multirow}
\usepackage{booktabs}
\usepackage{textcomp}
\usepackage{gensymb}
\usepackage{nomencl}
\usepackage{comment}
\usepackage{setspace}
\usepackage{geometry}
\usepackage[section]{placeins}
\usepackage{soul}
\journal{Elsevier}

\makeatletter
\def\ps@pprintTitle{%
	\let\@oddhead\@empty
	\let\@evenhead\@empty
	\def\@oddfoot{}%
	\let\@evenfoot\@oddfoot}
\makeatother

\begin{document}
\captionsetup[figure]{labelfont={bf},name={Fig.},labelsep=period}
\begin{frontmatter}

\title{Topological edge states of quasiperiodic elastic metasurfaces}


\author[add1]{Xingbo Pu}
\author[add1]{Antonio Palermo\corref{corr1}}
\ead{antonio.palermo6@unibo.it}
\author[add1]{Alessandro Marzani\corref{corr1}}
\ead{alessandro.marzani@unibo.it}
\cortext[corr1]{Corresponding authors}

\address[add1]{Department of Civil, Chemical, Environmental and Materials Engineering, University of Bologna, 40136 Bologna, Italy}

\begin{abstract}

In this work, we investigate the dynamic behavior and the topological properties of quasiperiodic elastic metasurfaces, namely arrays of mechanical oscillators arranged over the free surface of an elastic half-space according to a quasiperiodic spatial distribution. An ad-hoc multiple scattering formulation is developed to describe the dynamic interaction between Rayleigh waves and a generic array of surface resonators. The approach allows to calculate the spectrum of natural frequencies of the quasiperiodic metasurface which reveals a fractal distribution of the frequency gaps reminiscent of the Hofstadter butterfly. These gaps have nontrivial topological properties and can host Rayleigh-like edge modes. We demonstrate that such topologically protected edge modes can be driven from one boundary to the opposite of the array by a smooth variation of the phason, a parameter which modulates the geometry of the array. Topological elastic waveguides designed on these principles provide new opportunities in surface acoustic wave engineering for vibration control, energy harvesting, and lossless signal transport, among others.

\end{abstract}


\end{frontmatter}


\section{Introduction}

The discovery of topological insulators in condensed matter physics has fueled the research interest towards the design of materials and devices able to control the transport of energy in several branches of physics, including electromagnetism \cite{khanikaev2013photonic, ozawa2019topological}, acoustics \cite{yang2015topological, he2018topological, chenyafeng2021} and elasticity \cite{wang2015topological, hu2021deep, hu2022local, chen2022topology, wen2022topological}. The quest for topological waveguides stems from their ability to support the propagation of robust edge states which are immune to the presence of defects or imperfections. The propagation of such defect-immune interface states has been demonstrated in numerous two-dimensional (2D) domains, including examples of acoustic and elastic metamaterials able to replicate topological phenomena like Hall \cite{ma2019topological}, spin Hall \cite{he2016acoustic}, and the quantum valley Hall effects \cite{lu2017observation, huo2021experimental}. In the above examples, the existence of protected edge states inside nontrivial bulk band gaps originates from broken symmetries within the periodic systems, either in time or space.

A companion strategy to engineer topological states makes use of quasiperiodic structures. The most interesting property of quasiperiodicity relevant to topological phenomena is its relation to higher dimensions: a quasiperiodic function can be regarded as a slice of a periodic function of a higher dimension, the superspace \cite{kraus2016quasiperiodicity}. The topological properties of the superspace can manifest in its low-dimensional quasiperiodic counterpart. This idea has been confirmed by a number of experiments in both photonic \cite{kraus2012topological, zilberberg2018photonic, lohse2018exploring} and acoustic systems \cite{apigo2019observation, ni2019observation, chen2021creating}.

Recently, the interest in the topological properties of quasiperiodic systems has extended to the mechanical community. In particular, quasiperiodic structures for flexural waves in beams \cite{pal2019topological, xia2020topological, rosa2019edge, rosa2021exploring, riva2020adiabatic} and plates \cite{riva2020edge, Dani2021dipolar, Dani2021edge} have demonstrated the existence of topological gaps and edge states in Hofstadter-like spectra \cite{hofstadter1976energy}. Due to their ease in construction and flexibility in tuning, quasiperiodic elastic metamaterials provide simple pathways to realize novel devices for vibration isolation, energy harvesting, and wave propagation control.

Although the above-mentioned contributions advanced the knowledge of topological physics in mechanical systems, the design of quasiperiodic structures to control surface waves in a three-dimensional (3D) semi-infinite medium has so far not been achieved. Indeed, the control of surface waves (SAWs) using locally resonant metamaterials, aka metasurfaces, has shown promising applications ranging from vibration isolation \cite{colquitt2017seismic, pu2020seismic}, energy harvesting \cite{chaplain2020topological} and non-reciprocal signal propagation \cite{wu2021non, palermo2020surface}. In this context, the propagation of immune-to-defect SAWs offers promising opportunities in modern communication systems, where SAW filters are already used to process radio-frequency signals in portable communication devices. 
 
Hence, in this work we study how surface waves of the Rayleigh type interact with a quasiperiodic array of pillar-shaped resonators and demonstrate the existence of topological gaps and edge states in a 3D elastic half-space. To this purpose, we develop and leverage a 3D multiple scattering formulation to model the mutual interactions between resonators arranged in a generic cluster. Multiple scattering techniques are reliable tools to investigate the dynamics of aperiodic resonant systems, as recently shown for flexural waves in plates \cite{Dani2021dipolar, Dani2021edge} and surface waves in 2D elastic media \cite{PU2021lamb}.

Our quasiperiodic patterns result from a cyclic modulation of a periodic array (see Fig. \ref{fig:fig1}). By smoothly varying the modulation length, we obtain a family of periodic and quasiperiodic configurations which possess a Hofstadter-like spectrum. Then, using a topological invariant, the Chern number, we classify the topology of the band gaps found in the spectrum. Furthermore, we demonstrate the existence of edge modes and their localization in the array with respect to the phason parameter.

Our paper is organized as follows. Following this introduction, we present the problem statement in Section \ref{Statement of the problem}. In Section \ref{3D Multiple scattering formulation} we develop a 3D multiple scattering formulation and describe the  solution strategy to compute the frequency spectra. Section \ref{Topological band gaps and edge states} discusses the topological properties of the frequency spectra,  the existence of edge states and their location according to the phason parameter. Finally, we summarize the main conclusions and outlook of our work in Section \ref{Conclusion}.

\section{Statement of the problem} \label{Statement of the problem}

\begin{figure}[htbp]
	\centering
	\includegraphics[width=0.7\textwidth]{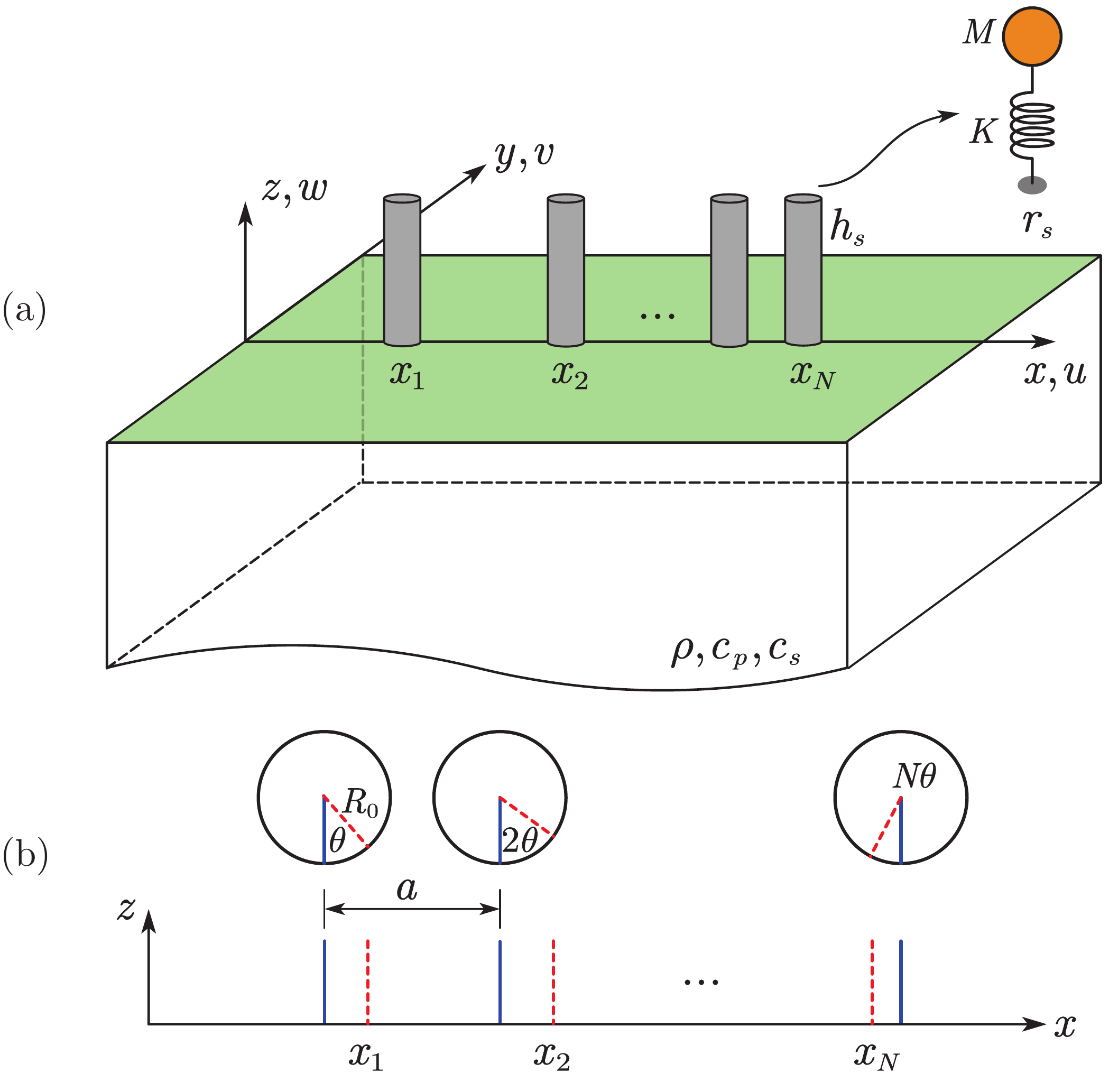}
	\caption{Schematic of a finite quasiperiodic array of resonators: (a) array of cylindrical pillars atop a 3D elastic half-space, (b) solid and dashed lines indicate the resonator locations before and after the modulation, respectively.}
	\label{fig:fig1}
\end{figure}

Let us consider a 3D elastic half-space coupled at the free surface with an array of $N$ elastic cylindrical pillars, which are distributed along the $x$ direction according to the family of periodic and quasiperiodic patterns \cite{apigo2018topological, pal2019topological}:

\begin{equation} \label{equ:space modulation expresion}
    x_n = na+R_0\sin(2\pi n\theta + 2\pi \phi), \quad n=1,...,N.
\end{equation}

\noindent where $x_n$ is the position of the $n$-th pillar, $a$ is the lattice constant, $R_0$ is the radius of the modulation circle,  $\theta$ is the  modulation period and $\phi \in [0, 1]$ the related phase (see Fig. \ref{fig:fig1}). For an infinite cluster of pillars $N=\infty$, the value of the parameter $\theta$  discriminates between periodic and quasiperiodic configurations. In particular, if $\theta=\gamma/\beta$ is a rational number, with $\gamma$ and $\beta$ being coprime integers, then the spatial periodicity of the pillars is $\beta a$. Conversely, if $\theta$ is a irrational number there is no translation symmetry along the array, thus resulting in a quasiperiodic configuration \cite{xia2020topological}. The phase $\phi$, also referred to as phason \cite{apigo2019observation}, does not affect the periodicity length of the infinite array while produces a cyclic modulation of the resonator locations \cite{pal2019topological}.

We focus our interest on arrays composed of identical pillars of radius $r_s$ and height $h_s$. To avoid any overlapping between two generic adjacent pillars, we constrain their footprint width (i.e., diameter)  to fulfill the inequality:

\begin{equation} 
     x_{n+1}-x_n=a+R_0[\sin(2\pi n\theta+2\pi \theta+2\pi \phi)-\sin(2\pi n\theta+2\pi \phi)]\ge a-2R_0\ge 2r_s, 
\end{equation}
\noindent
which yields $R_0\le (a-2r_s)/2$.

To describe the dynamic response of each pillar, we resort to a discrete single-degree-of-freedom model, with stiffness $K$ and mass $M$  (see the inset in Fig. \ref{fig:fig1}a), thus considering only its vertical motion. As a result, each resonator exchanges with the substrate a normal stress which is uniformly distributed over the resonator base  $S_n=\pi r_s^2$ \cite{PU2021lamb}.

For this configuration, we aim to discuss the existence and nature of localized surface modes of the Rayleigh type which rise from the collective response of a finite array of $N$ resonators excited by the substrate wave field. To model the mutual interaction between the resonators and the half-space, we develop and exploit a 3D multiple scattering formulation which is detailed in the next section.

\section{3D Multiple scattering formulation} \label{3D Multiple scattering formulation}

\subsection{Elastic wave field}
We formulate the elastodynamic governing equations in a 3D setting described by the spatial coordinates $(x,y,z)=(\mathbf{r},z)$, where $|\mathbf{r}|=\sqrt{x^2+y^2}$ is considered for the cylindrical reference system with the same origin as for $(x,y,z)$. The displacement components in the half-space are denoted by $u(\mathbf{r},z)$, $v(\mathbf{r},z)$ and $w(\mathbf{r},z)$ along $x$, $y$ and $z$ directions. We restrict our attention to time-harmonic regime and omit the related term $\mathrm{e}^{\mathrm{i} \omega t}$ through the rest of the derivation.

Assuming an incident wave field with displacement components $\mathbf{u}_0=[u_0,v_0,w_0]$ impinging the bases of the $N$ resonators, the total wave field at the generic position $(\mathbf{r},z)$ can be expressed as the summation of the incident and scattered wave fields of the $N$ resonators:

\begin{subequations}
\begin{equation} \label{equ:expression of total u}
     u(\mathbf{r},z)=u_0(\mathbf{r},z)+\sum_{n=1}^{N} Q_n G_u(\mathbf{r}-\mathbf{r}_n,z),
\end{equation}
\begin{equation} \label{equ:expression of total v}
     v(\mathbf{r},z)=v_0(\mathbf{r},z)+\sum_{n=1}^{N} Q_n G_v(\mathbf{r}-\mathbf{r}_n,z),
\end{equation}
\begin{equation} \label{equ:expression of total w}
     w(\mathbf{r},z)=w_0(\mathbf{r},z)+\sum_{n=1}^{N} Q_n G_w(\mathbf{r}-\mathbf{r}_n,z),
\end{equation}
\end{subequations}

\noindent where $Q_n$ is the amplitude of the uniform normal stress at the base of the $n$-th resonator, and $G_u$, $G_v$ and $G_w$ are the related displacement Green's functions, namely the displacement components along $x$, $y$ and $z$ directions.

The stress amplitude $Q_n$ is obtained from the dynamic equilibrium equation of the $n$-th resonator for an imposed vertical harmonic motion $w(\mathbf{r}_n,0)$ at its base:

\begin{equation} \label{equ:impedance Z}
     Q_n=\frac{M_n\omega_{rn}^2\omega^2}{S_n(\omega_{rn}^2-\omega^2)}w(\mathbf{r}_n,0)\equiv Z_n w(\mathbf{r}_n,0), \quad n=1,...,N,
\end{equation}

\noindent where $\omega_{rn}=\sqrt{K_n/M_n}$ is the resonator resonant frequency. Eq. \eqref{equ:impedance Z} indicates that the amplitude $Q_n$ is governed by both the resonator impedance $Z_n(M_n,S_n,\omega_{rn},\omega)$ and the total vertical displacement $w(\mathbf{r}_n,0)$ at its footprint position $(\mathbf{r}_n,0)$.

We set up the multiple scattering problem by using the wave dilatational  $\Phi(x,y,z)$ and distorsional $\Psi_j(x,y,z)$ $(j=x,y,z)$ potentials. For elastic waves in a 3D half-space, the potentials $\Phi(x,y,z)$ and $\Psi_j(x,y,z)$ satisfy the wave equations:
\begin{equation} \label{equ:3d wave equation}
    \nabla^{2} \Phi + k_p^{2} \Phi=0,  \quad  \nabla^{2} \Psi_j + k_s^{2} \Psi_j=0, \quad -\infty < x,y < \infty,z\le 0.
\end{equation}
\noindent where $k_p=\omega/c_p$ and $k_s=\omega/c_s$ denote the compressional and shear wavenumbers in the half-space, respectively, being $c_p$ and $c_s$ the compressional and shear wave velocities.

The Fourier transforms of Eqs. \eqref{equ:3d wave equation} along $x$ and $y$ directions read:

\begin{equation} \label{equ:3d transformed wave equation}
    \frac{\mathrm{d}^2 \bar{\Phi}}{\mathrm{d} z^2}-(k_x^2+k_y^2-k_p^2) \bar{\Phi}=0,  \quad  \frac{\mathrm{d}^2 \bar{\Psi}_j}{\mathrm{d} z^2}-(k_x^2+k_y^2-k_s^2) \bar{\Psi}_j=0,
\end{equation}
\noindent
in which $k_x$, $k_y$ denote the wavenumber in $x$ and $y$ directions, respectively. The general solutions of Eqs. \eqref{equ:3d transformed wave equation} have the form:

\begin{equation} \label{equ:transformed potentials}
    \bar{\Phi}=A \mathrm{e}^{pz}, \quad \bar{\Psi}_j=B_j \mathrm{e}^{qz},
\end{equation}
\noindent
with:
\begin{equation} \label{equ:wave numbers of potentials}
p=\sqrt{k_x^2+k_y^2-k_p^2}=\sqrt{k^2-k_p^2}, \quad q=\sqrt{k_x^2+k_y^2-k_s^2}=\sqrt{k^2-k_s^2},
\end{equation} 

\noindent and where the unknown coefficients $A$, $B_j$ can be determined by enforcing stress related boundary conditions. To this end, we  express the Cauchy stress components as functions of potentials via Hooke's law:

\begin{subequations}
	\begin{equation} \label{equ:sigma_zz}
	\sigma_{zz}(x,y,z)=\lambda \left(\frac{\partial^2}{\partial x^2} +\frac{\partial^2}{\partial y^2}+\frac{\partial^2}{\partial z^2}\right)\Phi+2\mu \left(\frac{\partial^2}{\partial z^2}\Phi -\frac{\partial^2}{\partial y \partial z}\Psi_x+\frac{\partial^2}{\partial x \partial z}\Psi_y\right), 
	\end{equation}
	\begin{equation} \label{equ:tau_zx}
	\tau_{zx}(x,y,z)=\mu \left[2\frac{\partial^2}{\partial x \partial z}\Phi -\frac{\partial^2}{\partial x \partial y}\Psi_x+\left(\frac{\partial^2}{\partial x^2}-\frac{\partial^2}{\partial z^2} \right)\Psi_y+\frac{\partial^2}{\partial y \partial z}\Psi_z\right],
	\end{equation}
	\begin{equation} \label{equ:tau_zy}
	\tau_{zy}(x,y,z)=\mu \left[2\frac{\partial^2}{\partial y \partial z}\Phi+\left(\frac{\partial^2}{\partial z^2}-\frac{\partial^2}{\partial y^2} \right)\Psi_x+\frac{\partial^2}{\partial x \partial y}\Psi_y -\frac{\partial^2}{\partial x \partial z}\Psi_z\right],
	\end{equation}
\end{subequations}
\noindent
in which $\lambda$ and $\mu$ are the Lam\'e constants. 
Then, we assume a uniformly distributed normal stress acting on the footprint of each resonator, i.e., over the surface area $S_n=\pi r_s^2$. The boundary conditions are thus expressed as:

\begin{equation} \label{equ:B.C for 3d problem}    
\sigma_{zz} (x,y,0) = \left\{
\begin{array}{rl}
1 & \text{if}\, \sqrt{x^2+y^2} \le r_s\\
0 & \text{elsewhere}
\end{array}, \right. \quad
\tau_{zx} (x,y,0) =0, \quad \tau_{zy} (x,y,0) = 0.
\end{equation}

Note that the potential $\mathbf{\Psi}$ is a divergence-free vector field that satisfies $\nabla \cdot \mathbf{\Psi}=0$. This additional constrain must be considered here, since the three equations in Eqs. (\ref{equ:sigma_zz}, \ref{equ:tau_zx}, \ref{equ:tau_zy}) contain four unknowns $A, B_j$ $(j=x,y,z)$ \cite{yang2009wave}. 

Fourier transforming Eqs. (\ref{equ:sigma_zz}, \ref{equ:tau_zx}, \ref{equ:tau_zy}, \ref{equ:B.C for 3d problem}) and considering $\nabla \cdot \mathbf{\Psi}=0$ yields:


\begin{subequations} 
\begin{equation} \label{equ:coefficient A}
    A=\frac{2k^2-k_s^2}{\mu R(k)}\cdot\frac{2\pi r_s}{k}J_1(kr_s),
\end{equation}
\begin{equation} \label{equ:coefficient B_x}
    B_x=\frac{-2\mathrm{i}p k_y}{\mu R(k)}\cdot\frac{2\pi r_s}{k}J_1(kr_s),
\end{equation}
\begin{equation} \label{equ:coefficient B_y}
    B_y=\frac{2\mathrm{i}p k_x}{\mu R(k)}\cdot\frac{2\pi r_s}{k}J_1(kr_s),
\end{equation}
\begin{equation} \label{equ:coefficient B_z}
    B_z=0,
\end{equation}
\end{subequations}
\noindent
where $J_1(\cdot)$ is the Bessel function of the first kind of order one and $R(k)$ is the Rayleigh function \cite{lamb1904propagation}:
\begin{equation} 
    R(k)=(2k^2-k_s^2)^2-4k^2 pq. 
\end{equation}

At last, by substituting Eqs. (\ref{equ:coefficient A}, \ref{equ:coefficient B_x}, \ref{equ:coefficient B_y}, \ref{equ:coefficient B_z}) into Eqs. \eqref{equ:transformed potentials} and by using inverse Fourier transform and Helmholtz decomposition, we obtain the sought Green's functions:

\begin{subequations}
\begin{equation} \label{equ:3d G_zu}
     G_u(x,y,z)=\frac{\mathrm{i}r_s}{2\pi \mu} \int_{-\infty}^{\infty}\int_{-\infty}^{\infty}\frac{k_x J_1(kr_s)[(2k^2-k_s^2)\mathrm{e}^{pz}-2pq\mathrm{e}^{qz}]}{kR(k)}\mathrm{e}^{\mathrm{i}(k_x x+k_y y)}\,\mathrm{d}k_x\mathrm{d}k_y, 
\end{equation}
\begin{equation} \label{equ:3d G_zv}
     G_v(x,y,z)=\frac{\mathrm{i}r_s}{2\pi \mu} \int_{-\infty}^{\infty}\int_{-\infty}^{\infty}\frac{k_y J_1(kr_s)[(2k^2-k_s^2)\mathrm{e}^{pz}-2pq\mathrm{e}^{qz}]}{kR(k)}\mathrm{e}^{\mathrm{i}(k_x x+k_y y)}\,\mathrm{d}k_x\mathrm{d}k_y, 
\end{equation}
\begin{equation} \label{equ:3d G_zw}
  G_w(x,y,z)=\frac{r_s}{2\pi \mu} \int_{-\infty}^{\infty}\int_{-\infty}^{\infty}\frac{p J_1(kr_s)[(2k^2-k_s^2)\mathrm{e}^{pz}-2k^2\mathrm{e}^{qz}]}{kR(k)}\mathrm{e}^{\mathrm{i}(k_x x+k_y y)}\,\mathrm{d}k_x\mathrm{d}k_y. 
\end{equation}
\end{subequations}

Eqs. (\ref{equ:3d G_zu}, \ref{equ:3d G_zv}, \ref{equ:3d G_zw}) can also be expressed with respect to a cylindrical coordinate system as \cite{miller1954field}:

\begin{subequations}
\begin{equation} \label{equ:3d Gr}
     G_{u_r}(\mathbf{r},z)=\frac{-r_s}{\mu} \int_{0}^{\infty}\frac{k J_1(kr_s)[(2k^2-k_s^2)\mathrm{e}^{pz}-2pq\mathrm{e}^{qz}]}{R(k)}J_1(k|\mathbf{r}|)\,\mathrm{d}k, 
\end{equation}

\begin{equation} \label{equ:3d Gw}
     G_w(\mathbf{r},z)=\frac{r_s}{\mu} \int_{0}^{\infty}\frac{p J_1(kr_s)[(2k^2-k_s^2)\mathrm{e}^{pz}-2k^2\mathrm{e}^{qz}]}{R(k)}J_0(k|\mathbf{r}|)\,\mathrm{d}k, 
\end{equation}
\end{subequations}
\noindent
where $J_0(\cdot)$ is the Bessel function of the first kind of order zero. The above Green's functions can be evaluated numerically via Gauss–Kronrod quadrature. To avoid numerical instabilities, we assume a small hysteretic damping ratio $\xi = 0.1\%$ in the substrate to remove the poles of the integrates \cite{PU2021lamb}.

\subsection{Solution strategy}

To obtain the coefficient $Q_n$, we substitute Eq. \eqref{equ:impedance Z} into Eq. \eqref{equ:expression of total w} and specify them at the resonator location $(\mathbf{r}_m,0)$:

\begin{equation} \label{equ:total w at xm}
	Z_m^{-1} Q_m = w_0(\mathbf{r}_m,0) +\sum_{n=1}^{N} Q_n G_w(\mathbf{r}_m-\mathbf{r}_n,0), \quad n, m=1,...,N.
\end{equation}

With some algebra,  Eq. \eqref{equ:total w at xm} can be reorganized in matrix form as:

\begin{equation}
    \mathbf{AX}=\mathbf{B}, \label{equ:Ax=b}
\end{equation}
\noindent
with:

\begin{equation}
    \mathbf{A} = \left[\begin{array}{cccc}
    {Z_1^{-1}-G_w(\mathbf{0},0)} & {-G_w(\mathbf{r}_1-\mathbf{r}_2,0)} & {\cdots} & {-G_w(\mathbf{r}_1-\mathbf{r}_N,0)} \\
    {-G_w(\mathbf{r}_2-\mathbf{r}_1,0)} & {Z_2^{-1}-G_w(\mathbf{0},0)} & {\cdots} & {-G_w(\mathbf{r}_2-\mathbf{r}_N,0)} \\
    {\vdots} & {\vdots} & {\ddots} & {\vdots} \\
    {-G_w(\mathbf{r}_N-\mathbf{r}_1,0)} & {-G_w(\mathbf{r}_N-\mathbf{r}_2,0)} & {\cdots} & {Z_N^{-1}-G_w(\mathbf{0},0)} \\
    \end{array}\right], \; \mathbf{X}=\left[\begin{array}{c}
    {Q_1} \\
    {Q_2} \\
    \vdots \\
    {Q_N} \\  
    \end{array}\right], \;
    \mathbf{B}=\left[\begin{array}{c}
    {w_0(\mathbf{r}_1,0)} \\
    {w_0(\mathbf{r}_2,0)} \\
    \vdots \\
    {w_0(\mathbf{r}_N,0)} \\  
    \end{array}\right].
\end{equation}

Thus, for a given incident wave field $w_0(\mathbf{r},0)$, the vector $\mathbf{X}$ of the stress amplitudes $Q_n$ can be computed as $\mathbf{X}=\mathbf{A}^{-1}\mathbf{B}$. Given the stress amplitudes $Q_n$, the total wave field is obtained using Eqs. (\ref{equ:expression of total u}, \ref{equ:expression of total v}, \ref{equ:expression of total w}).

Conversely, by assuming a null incident field (i.e., $w_0=0$), the multiple scattering formulation in Eq. \eqref{equ:total w at xm} degenerates to the eigenvalue problem:

\begin{equation}
    \mathbf{AX}=\mathbf{0}, \label{equ:Ax=0}
\end{equation}
\noindent
which provides the eigenstates of the system. A similar multiple scattering approach has been recently proposed in Ref. \cite{Dani2021edge} to discover the existence of flexural edge modes in quasiperiodic arrays of resonators over an elastic plate.

At this stage, the calculation of the nontrivial solutions of Eq. \eqref{equ:Ax=0} requires to identify those frequencies for which the determinant $|\mathbf{A}|$ is equal to zero; this is equivalent to search for a null eigenvalue of the matrix $\mathbf{A}$ for a given input frequency $\omega$. However, given the unbounded geometry of the system, only complex frequencies can meet this condition. Nonetheless,  following the approximation proposed in \cite{Dani2021dipolar, Dani2021edge}, we compute the minimum eigenvalue ($\lambda_{min}$) of $\mathbf{A}$ for given real frequency $\omega$, thus neglecting its imaginary component. For localized modes, this approximation yields negligible discrepancies. Thus, the eigenvalue problem can be expressed as:

\begin{equation} \label{equ:eigenvalue problem}
    \mathbf{A} \mathbf{X} = \lambda_{min} \mathbf{X}, \quad (\lambda_{min} \to 0).
\end{equation}
\noindent The resulting eigenvector $Q_n$ are used in Eqs. (\ref{equ:expression of total u}, \ref{equ:expression of total v}, \ref{equ:expression of total w}) to calculate the eigenfields in the half-space.

\section{Topological band gaps and edge states} \label{Topological band gaps and edge states}

In this section, we examine the dynamics of an array composed by $N=30$ identical resonators (see Fig. \ref{fig:fig1}) whose locations obey Eq. \eqref{equ:space modulation expresion}. Arrays of similar dimensions have been considered to identify topological edge states in elastic beams equipped with quasiperiodic resonators \cite{xia2020topological}. In what follows, we inspect the novel dynamic behavior of a family of configurations associated with different values of $\theta$ and $\phi$. The mechanical and geometrical parameters of the half-space and the resonators array are collected in Table \ref{tab:tab1}. For a mass-spring resonator of mass $M$, we introduce the mass ratio $\hat{m}=M/M_s$, which relates the resonator mass to a conventional substrate mass $M_s=\rho \lambda_r S$, where $\rho$ is the substrate density, $\lambda_r$ the Rayleigh wavelength at resonant frequency $\omega_r$ and $S=\pi r_s^2$ the resonator footprint area.

\subsection{Hofstadter butterfly spectrum} 

\begin{figure}[htbp]
	\centering
	\includegraphics[width=1\textwidth]{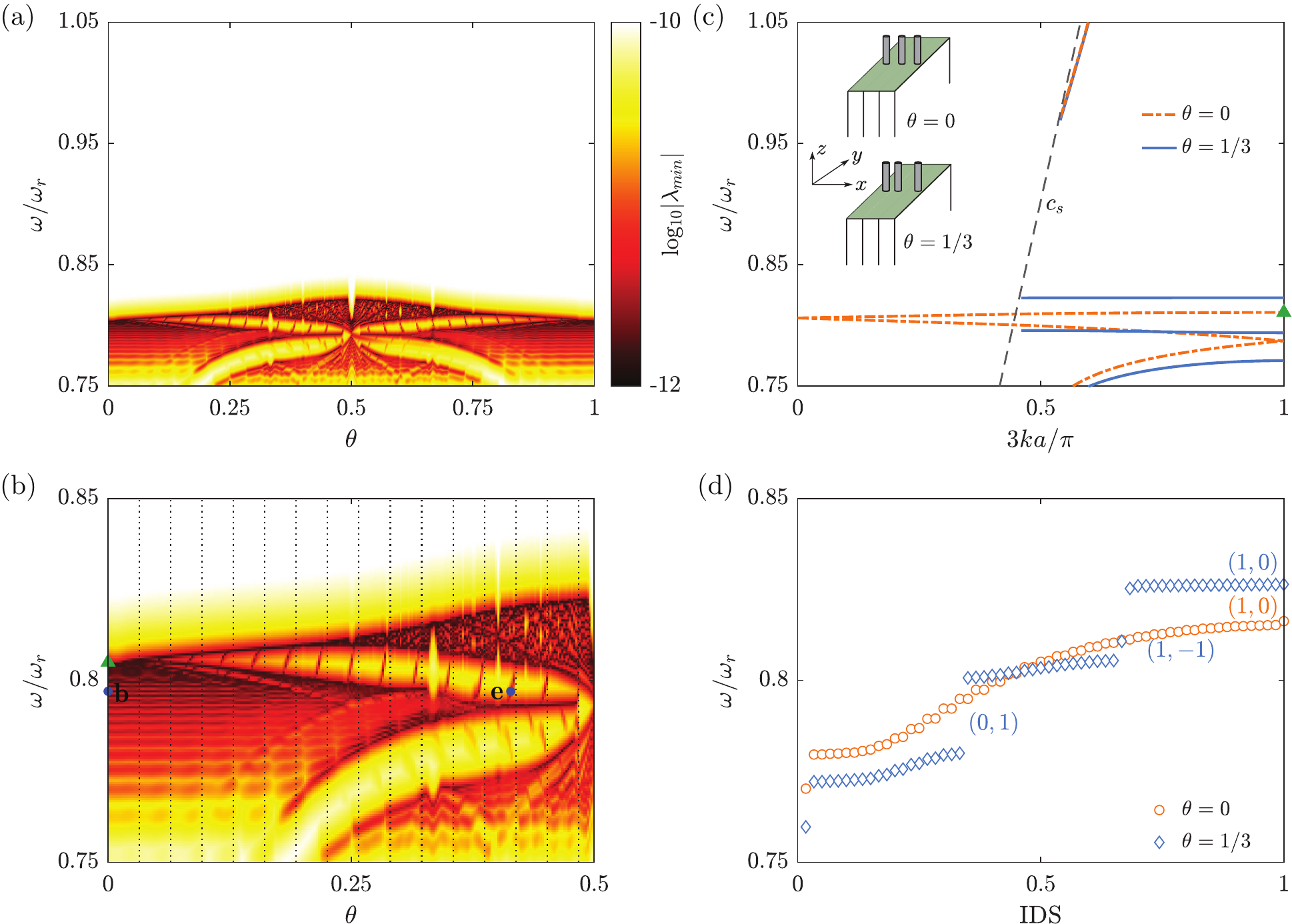}
	\caption{(a) Hofstadter butterfly spectrum of a finite array of resonators ($N=30$) atop a 3D elastic half-space with space modulation $\theta$ ($\phi=0$) in Eq. \eqref{equ:space modulation expresion}, in which each $\theta$ corresponds to a specific configuration. (b) Zoomed-in map of the spectrum. Black dashed lines represent commensurate values of $\theta$. (c) Rayleigh-wave dispersion curves in infinitely periodic systems for $\theta=0$ and $\theta=1/3$, respectively. (d) IDS for $\theta=0$ and $\theta=1/3$, together with their topological invariants $(n,m)$ for each band gap.}
	\label{fig:fig2}
\end{figure}

From Eq. \eqref{equ:eigenvalue problem}, we compute the  eigenvalue $\lambda_{min}$ of the matrix $\mathbf{A}$ for several resonator patterns obtained by varying  $\theta=[0,1]$ and imposing a $\phi=0$.  The colormap of  $\log_{10}{|\lambda_{min}|}$ in Fig. \ref{fig:fig2}a is obtained for an array with resonators of mass ratio $\hat{m}=2$ in the frequency range $[0.75,1.05]\omega_r$, namely where strong scattering effects are expected. Regions with the darkest colors mark the existence of eigensolutions. Conversely, the lightest colors connote regions with frequency gaps. Notably, the distribution of these regions with respect to the parameter $\theta$ is reminiscent of the Hofstadter butterfly \cite{hofstadter1976energy}, which is characterized by a peculiar fractal network of frequency gaps. The fractal structure lies below $\omega_r$ and is symmetric with respect to $\theta=0.5$, where the larger fractal gaps are folded. The location and extensions of these gaps largely vary with the tuning parameter $\theta$, as better shown in the zoomed-in map of Fig. \ref{fig:fig2}b. 

Additionally, the reader can appreciate the presence of eigensolutions crossing the fractal gaps. As we shall see later, these modes have a localized nature, i.e., they are edge modes, and distribute within the spectrum according to a peculiar pattern dictated by the size of the finite array.

The frequency bounds of the fractal gaps at rational values of $\theta=\gamma/\beta$ can be predicted from the dispersive properties of the related infinite periodic arrays, i.e., with lattice constant $\beta a$ \cite{pal2019topological}. As a representative example, we consider the pattern with $\theta=1/3$ and compute its dispersion curves using FE simulations in Comsol Multiphysics. To this purpose, we model a supercell of length $3a$ and distribute three resonators according to $\theta=1/3$. Additionally, we model the configuration with $\theta=0$, i.e., a periodic array with lattice constant $a$. The related  of length $3a$ with $\theta=0$ is used as a reference. The dispersion curves of Rayleigh waves propagating along these arrays are displayed in Fig. \ref{fig:fig2}c. The reader can refer to \cite{al2016guidance, zheng2021multiple} for the simulation details.

\begin{table}[htbp]
	\caption{Mechanical parameters for resonators and the elastic half-space.}
	\label{tab:tab1}
	\begin{tabular*}{\hsize}{@{}@{\extracolsep{\fill}}lll@{}}
		\toprule
		Symbol & Definition & Value  \\   
		\midrule
		$a$ & Lattice constant   & 0.1$\lambda_r$ \\
		$r_s$ & Radius of resonator footprint & 0.15$a$  \\
		$R_0$ & Radius of modulation circle & 0.3$a$  \\
		$\omega_r$ & Resonant frequency  & 100 rad/s  \\
		$\rho$ & Mass density of half-space  & 1200 kg/m$^3$ \\
		$c_p$ & Compressional wave velocity & 900 m/s \\
		$c_s$ & Shear wave velocity & 500 m/s \\
		$K_h$ & Equivalent half-space stiffness & 669 MPa \\
		$\xi$ & Hysteretic damping ratio & 0.1$\%$ \\
		\bottomrule
	\end{tabular*}
\end{table}

Considering the periodic configuration $\theta=0$, we observe the existence of a locally resonant band gap located well below $\omega_r$. In fact, since the half-space acts as a soft support for the resonators, the lower edge ($\omega_\ell$) of the locally resonant gap shift towards a frequency $\omega_\ell<\omega_r$. This frequency can be estimated in closed form as: 
\begin{equation}
    \omega_\ell = \omega_r \sqrt{\frac{K_h}{M \omega_r^2+K_h}}\approx 0.8\omega_r,
\end{equation}
\noindent
in which the equivalent half-space stiffness $K_h$ is calculated from Eq. \eqref{equ:3d Gw} by assuming the average force over the footprint area:

\begin{equation}
    K_h = \pi r_s^2/\bar{G}_w = \frac{-\pi r_s \mu}{2} \bigg/ \int_{0}^{\infty}\frac{k_s^2 p}{R(k)} \frac{[J_1(kr_s)]^2}{kr_s}\,\mathrm{d}k,
\end{equation}

\noindent and the value of $K_h$ can be found in Table \ref{tab:tab1}.
The upper edge of the gap ($\omega_u$) corresponds to the frequency value where the optical branch crosses the shear wave dispersion curve \cite{colquitt2017seismic}. We observe that within the locally resonant gap, no eigensolutions are found in the spectrum of the finite-size array (Fig. \ref{fig:fig2}a). Indeed, locally resonant gaps have trivial nature, hence cannot host  edge modes in finite structures \cite{xia2020topological}.  

Moving back to the dispersion properties of the supercell model, Fig. \ref{fig:fig2}c, we observe that the acoustic branch (dashed lines) for $\theta=0$ folds twice at the first Brillouin-zone boundary ($k=0$ and $k=\pi/3a$). Conversely, the acoustic branch for $\theta=1/3$ splits in multiple branches (solid lines) opening two additional frequency gaps below the resonant one. These two gaps correspond to the fractal gaps observed in the Hofstadter spectrum in Fig. \ref{fig:fig2}b at $\theta=1/3$. 

\subsection{Topological properties}

We now examine the topological properties of the fractal gaps. To this purpose, we compute the integrated density of states (IDS) for a given frequency $\omega$ \cite{apigo2018topological, pal2019topological}:
\begin{equation} 
    \mathrm{IDS}(\omega)=\lim _{N \rightarrow \infty} \frac{\sum_{i}\left[\omega_{i} \leqslant \omega\right]}{N},
\end{equation}
\noindent
in which $[\cdot]$ is 1 when the condition is satisfied and otherwise is 0. The IDS provides the number of eigenfrequencies below $\omega$ divided by the number $N$ of discrete resonators of the finite system, as the length is taken to infinity. According to \cite{apigo2018topological, prodan2019k}, the IDS of the generic band gap ($g$), namely at any $\omega_{g} \in [\omega_{\ell,g}, \omega_{u,g}]$, can be related to two integers and $\theta$ as \cite{apigo2018topological, prodan2019k}:

\begin{equation} \label{equ:IDS}
     n + m\theta=\mathrm{IDS}(\omega_g), \quad n, m \in \mathbb{Z}.
\end{equation}

In Eq. \eqref{equ:IDS} the integer $m$ is the first Chern number $C=\frac{\partial \mathrm{IDS}(\omega_g)}{\partial \theta}$, a topological invariant that is used to label a gap as trivial ($C=0$) or nontrivial ($C \ne 0$).

To label the gaps of the spectra in Fig. \ref{fig:fig2}a, we compute the IDS for a substrate supporting $N=60$ resonators. The finite array is modeled in Comsol Multiphysics, where the periodic boundary conditions are imposed to the left and right sides of the model. The computed IDS is shown in Fig. \ref{fig:fig2}d for $\theta=0$ and $\theta=1/3$. For the array with $\theta=0$, we observe that the first 60 surface modes lie on a single band, with the highest mode frequency corresponding to $\omega_\ell$ (the small discrepancy in frequency is due to the stiffer response of the FE model). The resonant gap is characterized by IDS = 1 and thus by a Chern number $C=0$, which identifies the resonant gap as trivial.

For $\theta=1/3$, the surface modes distribute along three bands and the two gaps among them correspond to the fractal gaps in Fig. \ref{fig:fig2}c. These additional gaps are characterized respectively by IDS = $\theta$ and IDS = $1-\theta$, corresponding to Chern numbers $C=1$ and $C=-1$, respectively. Nonzero values of the Chern number confirm that the fractal gaps are nontrivial. We remind that the absolute value of Chern number $|C|$ correspond to the number of edge modes spanning the nontrivial gap between two subsequent commensurate values of $\theta$. The commensurate values $\theta_j=\frac{j}{N+1}$ for the finite array of $N=30$ resonators are reported in Fig. \ref{fig:fig2}b as dashed lines. In addition, a negative values of $C$ indicate that the related edge mode crosses the gap from the bottom to the top edge  with increasing $\theta$, while a positive $C$ indicates the opposite phenomenon \cite{pal2019topological, xia2020topological}. As an example, the reader can refer to the edge mode ``e" marked with a blue dot in Fig. \ref{fig:fig2}b which is enclosed within the commensurate values $\theta=12/31$, $\theta=13/31$ and is characterized by $C=-1$.

\begin{figure}[htbp]
	\centering
	\includegraphics[width=1\textwidth]{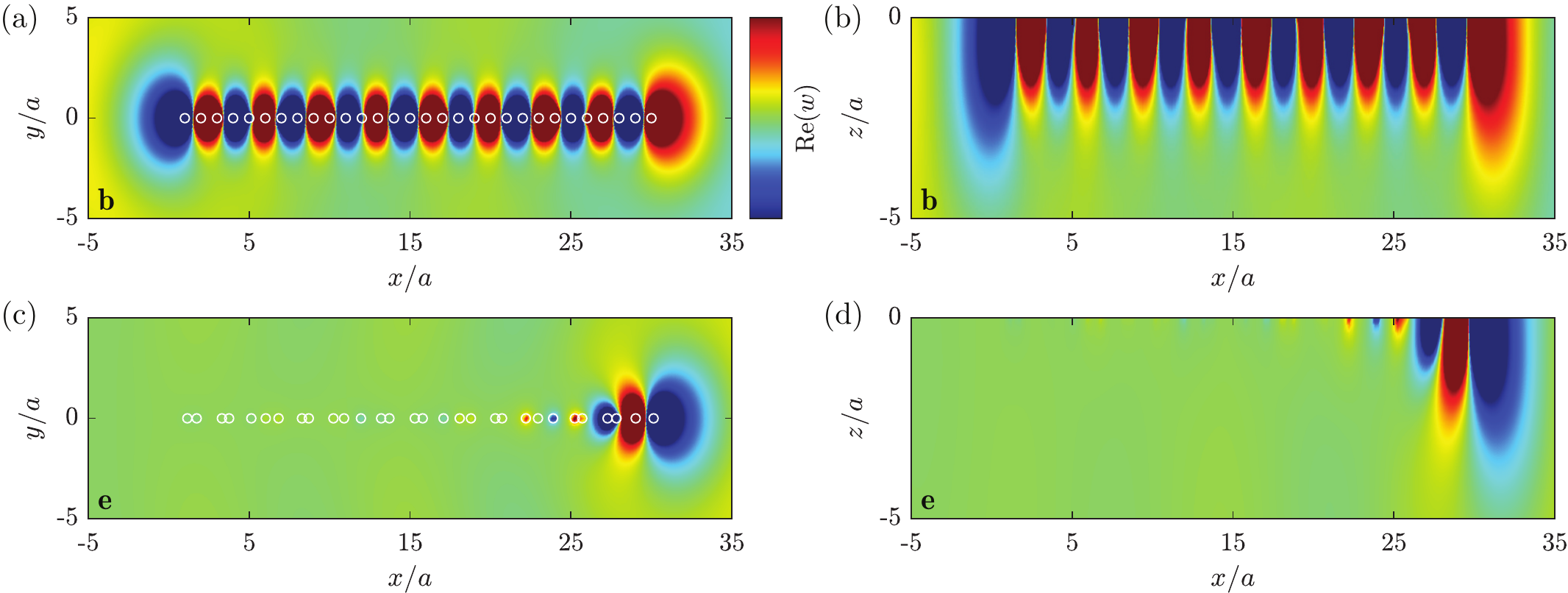}
	\caption{Representative eigenmodes at $\omega=0.797\omega_r$ labeled by ``b" ($\theta=0$), and ``e" ($\theta=\sqrt{2}-1$) in Fig. \ref{fig:fig2}b. (a), (b) Mode ``b" in $x-y$, and $x-z$ plane, respectively. (c), (d)  Mode ``e" in $x-y$, and $x-z$ plane, respectively. The resonator locations are labeled by white circles in (a) and (c).}
	\label{fig:fig3}
\end{figure}

To confirm our description of the Hofstadter spectrum in Fig. \ref{fig:fig2}a, we consider two representative eigensolutions at the same frequency $\omega=0.797\omega_r$ but with different values of $\theta$,  labeled by ``b" ($\theta=0$) and ``e" ($\theta=\sqrt{2}-1$) in Fig. \ref{fig:fig2}b, respectively. We compute and plot the corresponding eigenfields by using Eqs. (\ref{equ:expression of total w}, \ref{equ:3d Gw}) in the domain $x=[-5,35]a$, $y=[-5,5]a$, $z=[-5,0]a$. Fig. \ref{fig:fig3}a and Fig. \ref{fig:fig3}b show the top ($x$, $y$) and section ($x$, $z$) views of the bulk mode ``b", where the colormap displays the real part of the vertical displacement components, i.e., $\operatorname{Re}(w)$. Such a mode, also called Rayleigh-Bloch mode, is localized inside the whole array and decays perpendicular to it \cite{porter2005embedded}.  

Conversely, for the edge mode ``e", we observe a strong localization at the right boundary of the quasiperiodic array, as shown in Fig. \ref{fig:fig3}c and Fig. \ref{fig:fig3}d. In \ref{Appendix}, the reader can appreciate how such localized modes can be excited by a vertical source located close to the edge of the array.

Finally, we remark that all the edge modes spanning nontrivial gaps are localized at the right boundary of the array, where the finite array presents a truncation w.r.t. its infinite counterpart \cite{pal2019topological}. In what follows, we discuss how the localization of these modes can be controlled by smoothly varying the phason $\phi$.

\subsection{Edge-bulk-edge transitions driven by phason modulations}

A variation of the phason $\phi$ parameter within the range $[0,1]$ produces cyclic modulation of the finite-array pattern. As recently shown in acoustics and for flexural waves in mechanics, a smooth modulation of the phason can be exploited to ``transport" a localized edge mode across the array \cite{apigo2019observation, ni2019observation, pal2019topological, rosa2019edge, riva2020adiabatic, riva2020edge, xia2021experimental}. In this section, we aim at demonstrating this phenomenon in our context, i.e., the transition from right-localized to left-localized Rayleigh-like edge modes. 

\begin{figure}[htbp]
	\centering
	\includegraphics[width=1\textwidth]{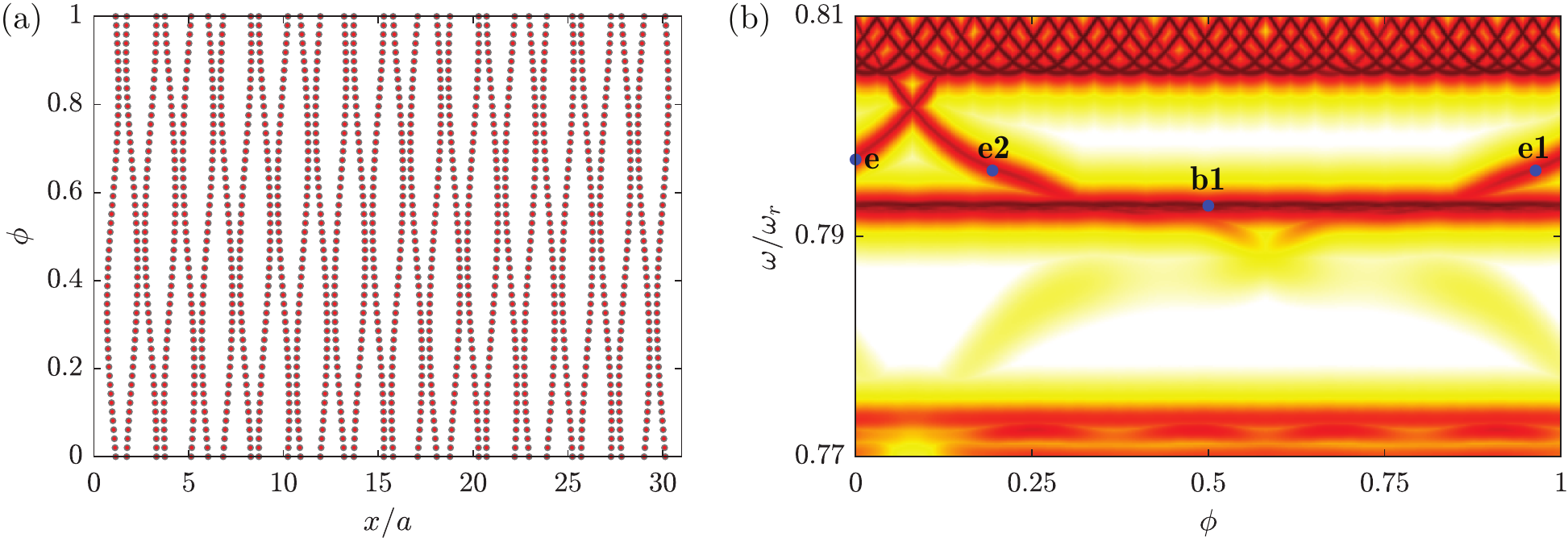}
	\caption{Phason modulation for finite resonators ($N=30$) with $\theta=\sqrt{2}-1$. (a) Location of resonators against $\phi$. (b) Spectra as a function of $\phi$.}
	\label{fig:fig4}
\end{figure}

To this purpose, we again consider the finite array with $\theta=\sqrt{2}-1$ and keep the remaining parameters unchanged. Fig. \ref{fig:fig4}a shows all the patterns of resonators as a function of $\phi$ in $[0,1]$. For such configurations, we compute the map of the minimum eigenvalue by varying $\omega$ within the fractal gaps, as displayed in Fig. \ref{fig:fig4}b. As expected, we observe some edge modes spanning the nontrivial gaps ($[0.774,0.792]\omega_r$ and $[0.794,0.805]\omega_r$) in the Hofstadter butterfly of Fig. \ref{fig:fig2}b. 

\begin{figure}[htbp]
	\centering
	\includegraphics[width=0.85\textwidth]{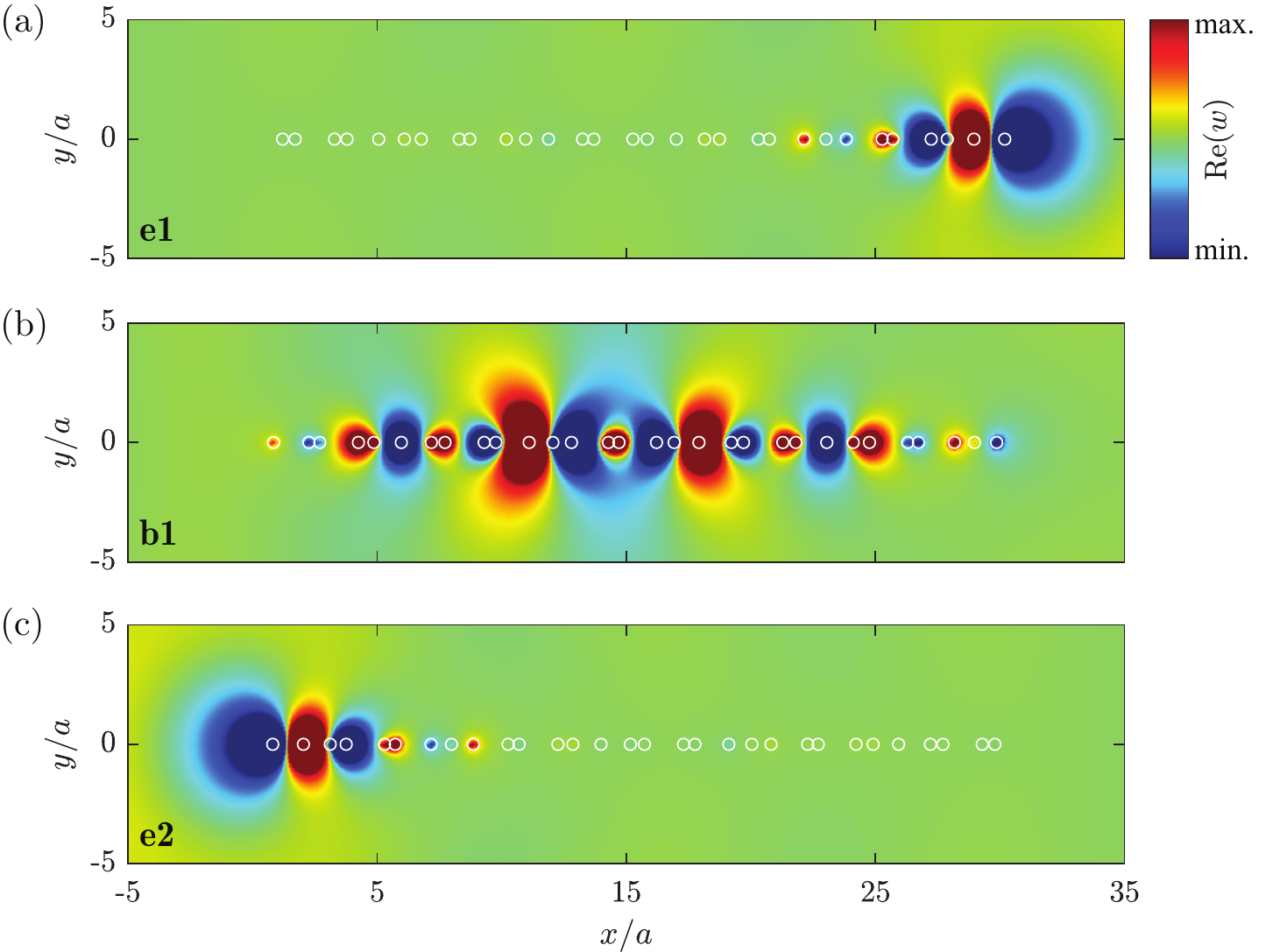}
	\caption{Representative eigenmodes label by ``e1", ``b1" and ``e2" in Fig. \ref{fig:fig4}b for: (a) $\phi=0.964, \omega=0.796\omega_r$, (b) $\phi=0.5, \omega=0.7928\omega_r$ and (c) $\phi=0.194, \omega=0.796\omega_r$. The resonator locations are labeled by white circles.}
	\label{fig:fig5}
\end{figure}

To examine the position of these modes within the array, we plot the wave fields of three representative eigenmodes labeled by ``e1" ($\phi=0.964, \omega=0.796\omega_r$), ``b1" ($\phi=0.5, \omega=0.7928\omega_r$) and ``e2" ($\phi=0.194, \omega=0.796\omega_r$), shown in Fig. \ref{fig:fig4}b and computed via Eqs. (\ref{equ:expression of total w}, \ref{equ:3d Gw}). The mode ``e1" is localized at the right boundary of the cluster (see Fig. \ref{fig:fig5}a), a region with low density of resonators (Fig. \ref{fig:fig4}a). Moving back to the spectra in Fig. \ref{fig:fig4}b, we observe that as the phason decreases, the edge mode gradually approaches the bulk band. When the branch comes in contact with the bulk band, its mode shape extends within the whole array (see ``b1" in Fig. \ref{fig:fig5}b). Finally, for smaller values of  $\phi$, the mode gets localized at the left boundary (see ``e2" in Fig. \ref{fig:fig5}c). We remark that the resonator location is 1-periodic with $\phi$, which means that the edge mode at $\phi=0$ (``e") is equivalent to the one at $\phi=1$. This ensures that the branch that crosses with ``e2" shares the same property as ``e1", the right-localized edge mode.

Overall, Fig. \ref{fig:fig5} illustrates the smooth evolution of the mode shape driven by a variation of the phason parameter $\phi$, which confirms the transition of localized Rayleigh-like states from right boundary, to the bulk, and finally to the left boundary. We anticipate that such a transition can be exploited to obtain topological pumping of Rayleigh-like edge modes.

\section{Conclusion} \label{Conclusion}

We have investigated the collective dynamics and topological properties of a cluster of mechanical resonators on a 3D half-space. By smoothly varying the parameter defining the resonator location, periodic and quasiperiodic patterns are generated, and the associated frequency spectra are computed using an ad-hoc developed multiple scattering formulation. The frequency spectra, computed as a function of the modulation parameter $\theta$, replicate the well-known Hofstadter butterfly, suggesting the existence of nontrivial gaps and edge modes for surface waves interacting with mechanical resonators. We have analyzed the topological properties of the fractal gaps of the Hofstadter-like spectrum by computing the integrated density of states and the related topological invariants. We have then demonstrated the existence of Rayleigh-like edge modes spanning the nontrivial gaps of a finite cluster of resonators. We have shown that these edge modes can be transferred from one boundary to the opposite boundary of the array by tuning the phason parameter. Our findings can serve as guidelines for future experiments on the localization of  surface edge modes and thus open a pathway for designing and realizing devices for wave localization, vibration mitigation and energy harvesting.

\section*{CRediT authorship contribution statement}
\noindent \textbf{Xingbo Pu:} Conceptualization, Methodology, Investigation, Software, Writing - original draft. \textbf{Antonio Palermo:} Conceptualization, Investigation, Validation, Writing - review \& editing, Supervision. \textbf{Alessandro Marzani:} Conceptualization, Investigation, Writing - review \& editing, Supervision, Funding acquisition.

\section*{Declaration of competing interest}
\noindent The authors declare that they have no conflict of interest.

\section*{Acknowledgments}
\noindent This project has received funding from the European Union’s Horizon 2020 research and innovation programme under the Marie Skłodowska Curie grant agreement No 813424. 

\appendix
\section{Harmonic excitation of edge modes}
\label{Appendix}
To demonstrate the excitability of the edge modes, we consider a vertical harmonic source, with frequency  $\omega=0.797\omega_r$, and distributed over a circular region of radius $r_s$, located on the half-space surface at $x=31a, y=0$. We compute the corresponding wave field in the domain $x=[-5,35]a$, $y=[-5,5]a$ by using Eq. \eqref{equ:expression of total w}. As a reference, we first provide the free field in Fig. \ref{fig:figA1}a, which clearly shows the free propagation and radiation of elastic waves. As expected, for $\theta=0$ and $\theta=\sqrt{2}-1$, the reader can appreciate that both bulk and edge modes can be excited and well captured by harmonic simulations, as clearly shown in Fig. \ref{fig:figA1}b and Fig. \ref{fig:figA1}c.

\setcounter{figure}{0}
\begin{figure}[hbt!]
	\centering
	\includegraphics[width=0.85\textwidth]{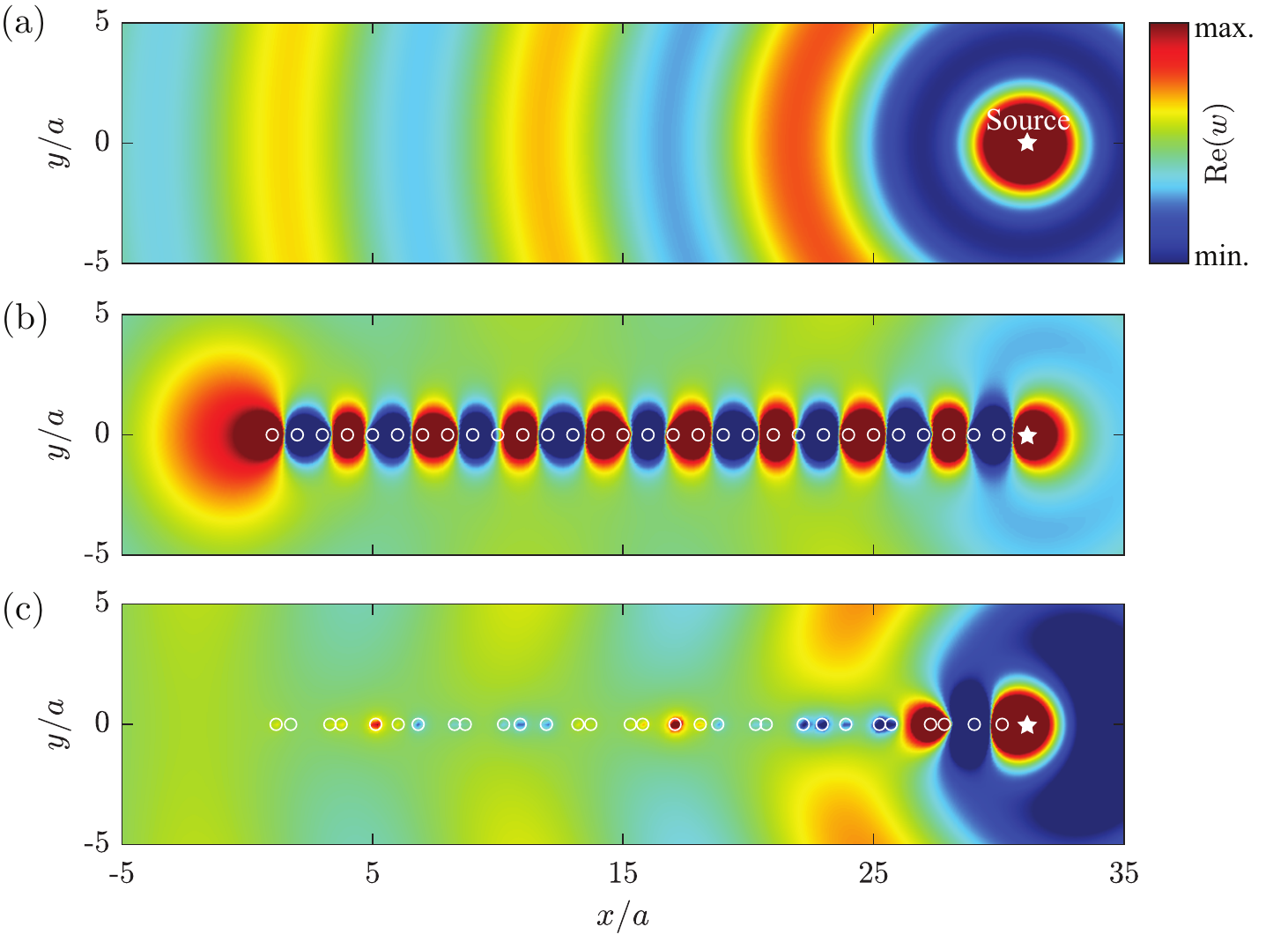}
	\caption{Harmonic wave fields on the half-space surface for: (a) free field, (b) $\theta=0$ and (c) $\theta=\sqrt{2}-1$. The incident wave at $\omega=0.797\omega_r$ is excited by a vertical distributed source marked as the star ($x=31a, y=0, z=0$). Resonator locations are labeled by white circles.}
	\label{fig:figA1}
\end{figure}

\bibliography{mybibfile}

\end{document}